\documentclass[twocolumn,showpacs,preprintnumbers,amsmath,amssymb,prl]{revtex4}
\usepackage{amsmath}
\usepackage{natbib}
\usepackage{epsfig}
\usepackage{amsmath,amssymb,mathrsfs}

\begin{document}

\title{Limits of slow-light in photonic crystals}

\author{Jesper Goor Pedersen, Sanshui Xiao, and Niels Asger Mortensen\footnote{email: asger@mailaps.org}}
\affiliation{DTU Fotonik, Department of Photonics Engineering, Technical University of Denmark,
Building 343, DK-2800 Kongens Lyngby, Denmark}

\date{\today}

\begin{abstract}
While ideal photonic crystals would support modes with a vanishing group velocity,
state-of-the art structures have still only provided a slow-down by roughly
two orders of magnitude. We find that the induced density of states
caused by lifetime broadening of the electromagnetic modes results in the group
velocity acquiring a finite value above zero at the band gap edges, while
attaining superluminal values within the band gap. Simple scalings of
the minimum and maximum group velocities with the imaginary part of the dielectric
function or, equivalently, the linewidth of the broadened states,
are presented. The results obtained are entirely general and may
be applied to any effect which results in a broadening of the electromagnetic
states, such as loss, disorder, finite-size effects, etc. This significantly
limits the reduction in group velocity attainable via photonic crystals.
\end{abstract}


\maketitle

The interest in slow-light phenomena has increased significantly in recent
years, due to the potential applications in areas such as
optical processing
\cite{opticalprocessing1,opticalprocessing2,opticalprocessing3,opticalprocessing4},
quantum information processing~\cite{quantinf1,quantinf2},
enhanced spontaneous emission~\cite{Lodahl2004},
and sensing~\cite{Mortensen2007, Pedersen2007}.
Strongly dispersive periodic structures with dielectric functions that vary
on the length scale of the wavelength of light can now be fabricated with
impressive precision and resolution~\cite{fabrication1,fabrication2}.
By exploiting the close analogies with condensed matter theory, these so-called
photonic crystals can give rise to such novel optical phenomena as e.g., enhanced
Raman scattering~\cite{McMillan2006}, increased stimulated emission~\cite{Lodahl2004},
superprism behavior~\cite{Kosaka1998}, and negative refraction~\cite{Notomi2000}.
Due to their highly dispersive
nature, photonic crystals have emerged as excellent candidates for sources of
slow-light phenomena~\cite{Notomi2001, Vlasov2005, Gersen2005, Altug2005}.
However, even though ideal structures would in principle support modes of vanishing
group velocity, state-of-the art structures have still only provided a slow-down
by roughly two orders of magnitude~\cite{Notomi2001, Altug2005}.
The limits imposed on the minimum attainable group velocity in photonic crystals
have been studied in various contexts, such as, e.g., fabrication
disorder~\cite{disorder1,disorder2}, lossy dielectrics~\cite{loss}, and
finite-size effects~\cite{finitesize}.
It is the aim of this paper to generalize
these findings and to show that they may all be presented in the context
of broadening of electromagnetic modes and the resulting induced density of states.

The existence of photonic band gaps in photonic crystals is
accompanied by van Hove singularities in the density of states (DOS)
near the band gap edge, which results in a vanishing group velocity.
In 1D the group velocity is inversely proportional to the density of
states, while in higher-dimensional structures it is inversely
proportional to the projected one-dimensional density of states
along the propagation direction, as in mesoscopic electron
transport. A formal proof can be given in terms of Wigner--Smith
group delay formalism, thus also applying to inhomogeneous
structures of finite extension (See e.g. Ref.~\cite{Brouwer1997} and
references therein).
Letting $g_0$ denote
the one-dimensional projected density of states along the propagation direction
of a homogenous material with a dielectric function given
as the average value of the dielectric function of the photonic crystal, we
identify three different regimes of interest:
\emph{(i)} a long wavelength regime, where the properties of the photonic crystal
are independent of the detailed geometrical composition, and thus $g(\omega)\simeq g_0$,
\emph{(ii)} a slow-light regime, where $g(\omega)>g_0$, and
\emph{(iii)} a "superluminal" regime, where $g(\omega)<g_0$.
We stress that $g(\omega)$ in all cases refers to the one-dimensional
projected density of states along the propagation direction, and that indeed
this is the case whenever we speak of the density of states.
Experiments have confirmed the existence of both
slow~\cite{Notomi2001, Vlasov2005, Gersen2005, Jacobsen2005}
and superluminal~\cite{Spielmann1994} regimes.

Consider the dispersion relation near the edge of a photonic band gap.
The exact shape of the dispersion relation
naturally depends on the geometry and dimensionality of the photonic
crystal. However, common to all band structures is a vanishing slope at the
band gap edge. If we expand the band structure near the
band gap edge $\omega(\mathrm{K})$$=$$\omega_0$ we thus obtain
\begin{equation}
\omega(k)\simeq\omega_0+\alpha(k-\mathrm{K})^2,
\end{equation}
where $\alpha$ is the group velocity dispersion (GVD).
Here we have ignored any higher-order contributions to the expansion.
Higher-order dispersion is of vital importance for the propagation
of pulses in the slow-light regime~\cite{Engelen2006}, but we ignore
it for now as we are only interested in studying the actual value of
the group velocity.
The group velocity becomes
\begin{equation}
v_g=\mathrm{Re}\left(\frac{\partial\omega}{\partial
k}\right)=
\mathrm{Re}\left(2\sqrt{\alpha(\omega-\omega_0)}\right), \label{eq:vg0}
\end{equation}
illustrating the square-root divergence of the density of states at
the band gap edge. Let us consider the effect of a finite imaginary
part of the dielectric function $\epsilon$$=$$\epsilon'+i\epsilon''$
of either of the constituents of the photonic crystal. To capture
the effect of a small imaginary part we apply standard
electromagnetic perturbation theory as described in detail in
Ref.~\onlinecite{Mortensen2006}.
The first order term in a perturbative expansion in the imaginary
part of the dielectric function is
\begin{equation}
\Delta\omega = -\frac{\omega}{2}
\frac{\left<\mathbf{E}\right| i\epsilon'' \left|\mathbf{E}\right>_{\mathcal{V}_1}}
{\left<\mathbf{E}\right|\epsilon'\left|\mathbf{E}\right>},
\end{equation}
where the integral in the denominator is restricted to the region
$\mathcal{V}_1$ of the constituent containing the imaginary dielectric
part. The effect is then an imaginary shift of the frequency
\begin{equation}
\Delta\omega=-\frac{1}{2}if\omega\epsilon''/\epsilon',
\end{equation}
where
\begin{equation}
f = \frac{\left<\mathbf{E}\right|\epsilon'
\left|\mathbf{E}\right>_{\mathcal{V}_1}}
{\left<\mathbf{E}\right|\epsilon'\left|\mathbf{E}\right>},
\end{equation}
is the fraction of dielectric energy localized in the corresponding
dielectric \cite{Joannopoulos1995}.
Making the substitution $\omega_0\rightarrow\omega_0+\Delta\omega_0$ in
Eq.~(\ref{eq:vg0}) this means that at the band edge
the group velocity becomes
%
\begin{equation}
v_g=\sqrt{\alpha f\omega_0\frac{\epsilon''}{\epsilon'}}.\label{eq:vg}
\end{equation}
Despite its simplicity, this results is very general and independent
of the particular photonic crystal geometry, be it of one, two
or three-dimensional nature. The result indicates that the
vanishing group velocity and the corresponding divergence of the
density of states is resolved in the case of a finite imaginary part
of the dielectric function. Furthermore, the group velocity has a
sub-linear dependence on $\epsilon''$, so the reduction in the group
index is significant, even for small imaginary parts of the
dielectric function.
In deriving this result we have of course assumed that the slow-down is
achieved at the band edge, where ideally $v_g=0$. Other schemes exist
for achieving slow-light within a flattened band, wherein the group
velocity is non-zero \cite{Ibanescu2004}. However, including such a term in the
previous derivation merely adds an additional term to Eq.~(\ref{eq:vg})
inside the square root,
and as such simply serves to increase the lower limit of the attainable
group velocity.
Due to the effects on pulse broadening, much attention has been
devoted to finding structures with low group velocity dispersion
in the slow-light regime \cite{Mori2005,Frandsen2006}.
Interestingly, our analysis reveals that such structures have
the additional benefit of reducing the minimum attainable
group velocity. Indeed, in the limit of vanishing GVD our analysis
shows that the group velocity attains the ideal value of zero,
provided that any higher-order dispersion is negligible.
Alternatively, we may consider the effect of
$\epsilon''$ as a finite broadening of the order $\epsilon''\omega$
of the electromagnetic modes~\cite{Mortensen2006}. Thus, any effect
causing such lifetime broadening of the modes is subject to a
similar analysis as just described, and will thus increase the
minimum group velocity as the square-root of the finite linewidth of
the broadened modes. This very general analysis applies to any
photonic crystal geometry, and any effect of lifetime broadening,
such as e.g., loss, disorder, or finite-size effects.
Of course, the analysis is based on a perturbative approach and as
such is only applicable in the case of small perturbations of a
perfect photonic crystal. In the limit of strong disorder, for example,
Anderson location will significantly alter the delay time statistics
of the photonic crystal~\cite{anderson1,anderson2}. In this analysis we limit ourselves to the
case where the mean-free path is sufficiently long so that the
concept of group velocity is meaningful.
In passing, we note that another effect of the broadening of the modes
is that a density of states is induced within the band gap. The induced
density of states mid-gap is
$g_{\mathrm{PBG}}\propto\epsilon''/(\Delta\omega)^2$, where
$\Delta\omega$ is the width of the gap~\cite{Mortensen2006}. The
result of this is that superluminal group velocities are attained
within the band gap, wherein the perturbative analysis reveals that
$v_g\propto1/\epsilon''$.

In the following we support these general findings by two
illustrative examples, where the source of the broadening is
a finite conductivity of the dielectric solid of an air-dielectric structure.
We first consider the simple case of a Bragg stack
with layers of width $a_1$$=$$0.8\Lambda$ and $a_2$$=$$0.2\Lambda$
for air and dielectric, respectively, where $\Lambda$ is the lattice constant.
We take $\epsilon_1$$=$$1$ for air, and assume that the dielectric
solid can be described by a frequency independent real part
$\epsilon'_2$$=$$9$ of the relative dielectric function.
The imaginary part is modeled as
$\epsilon''_2$$=$$\sigma/(\epsilon_0\omega)$, where $\sigma$ is
the conductivity of the dielectric and $\epsilon_0$ is the vacuum
permittivity. This is similar to the Drude model for the optical
response of metals, and is chosen so that the Kramers--Kronig
relations are fulfilled.

\begin{figure}
\begin{center}
\epsfig{file=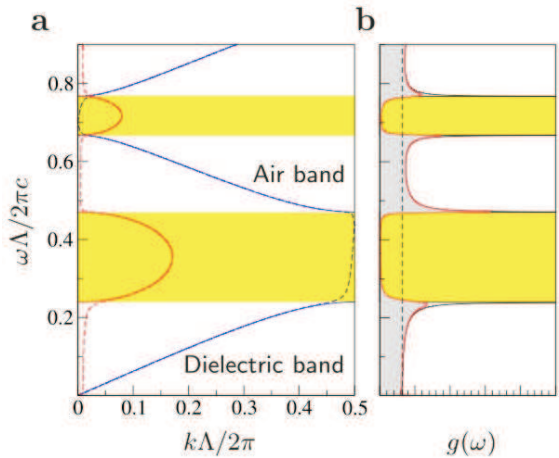, width=\linewidth,clip}
\caption{
(a) Photonic band structure of the Bragg stack. The
real parts of the $k$-vectors are shown in blue, while the
imaginary parts are shown in red. The dashed lines indicate
the case where the conductivity of the dielectric solid
is nonzero. The photonic band gaps are indicated by yellow
shading. (b), The corresponding density of states, with
the case of nonzero conductivity shown in red. The dashed horizontal
line indicates the density of states of a homogenous material with
$\epsilon=(a_1\epsilon_1+a_2\epsilon'_2)/\Lambda$.
}
\label{fig:braggBand}
\end{center}
\end{figure}
In Fig.~\ref{fig:braggBand} we show the band structure and corresponding density of
states of the Bragg stack for the two cases of zero and nonzero
conductivity of the dielectric solid, calculated using exact
analytical expressions~\cite{Kim2006}. At zero conductivity we note
that two photonic band gaps exist, where there are no states with
finite real $k$-values. These PBGs are caused by the complete
destructive interference of transmitted and reflected waves at
each layer of the Bragg stack. The loss caused by the nonzero
conductivity makes the destructive interference incomplete, and
the PBGs are no longer fully developed.
Consequently, a density of
states is induced in the band gap, and the divergence of the density
of states at the band gap edge is resolved.
The effect of this is two-fold.
Firstly, the group velocity at the band edge acquires a finite value larger
than zero because now $g$$\neq$$\infty$, and secondly, the group velocity
inside the PBG attains superluminal values larger than $c$ because in this
region $g$$<$$g_0$.
We stress that while the group velocity in the superluminal regime may exceed the
speed of light in vacuum, this implies no loss of causality, but is rather due
to a strong reshaping of the optical pulse,
as discussed for example by B\"uttiker and Washburn~\cite{Buttiker2003}.

\begin{figure}
\begin{center}
\epsfig{file=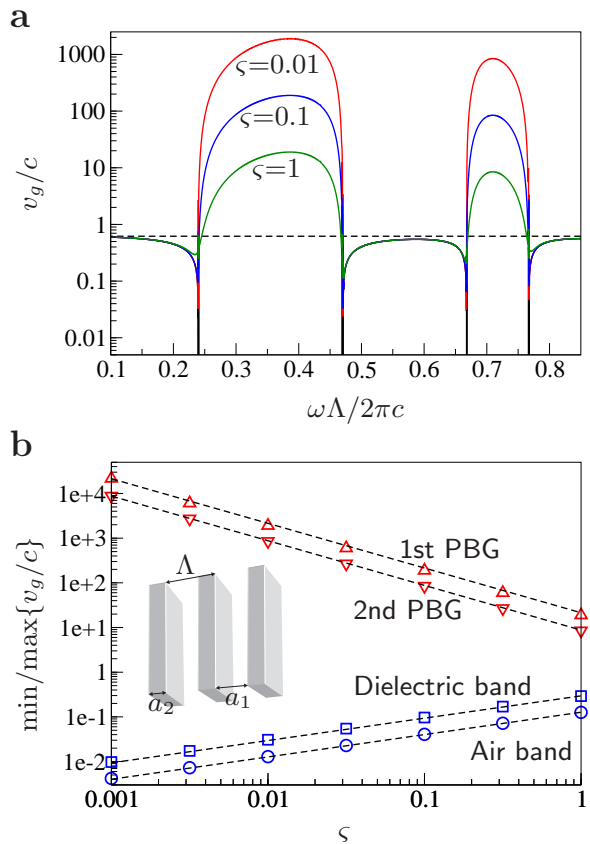, width=.9\linewidth,clip}
\caption{
(a) Group velocity as a function of
frequency for three different values of the dimensionless
parameter $\varsigma$=$\sigma\Lambda/(2\pi c\epsilon_0)$,
characterizing the conductivity of the dielectric solid. The solid
black line corresponds to the case of zero conductivity.
The dashed horizontal
line indicates the group velocity of a homogenous material with
$\epsilon=(a_1\epsilon_1+a_2\epsilon'_2)/\Lambda$.
(b) The minimum group velocities of the
dielectric and air bands of the first PBG, and the
maximum group velocities of the first and second PBG. All
are shown as a function of the dimensionless conductivity.
The dashed black lines indicate fits to the numerical
data of the form $A\sqrt{\varsigma}$ for the minimum group
velocities, and of the form $A/\varsigma$ for the maximum
group velocities. The inset shows the geometry of
the Bragg stack.
}
\label{fig:braggVg}
\end{center}
\end{figure}
In Fig.~\ref{fig:braggVg}a we show the group velocity as a function of
the normalized frequency, for three different values of the dimensionless
parameter $\varsigma$$=$$\sigma\Lambda/(2\pi c\epsilon_0)$, characterizing the
loss of the dielectric solid.
To calculate the group velocities we use the definition
$v_g=\partial\omega/\partial(\mathrm{Re}[k])$, which yields the same results as the formal
definition $v_g=\mathrm{Re}(\partial\omega/\partial k)$~\cite{Jackson1999}.
Even for a low value of $\varsigma$$=$$0.01$, we find that the
group velocity no longer drops below approximately 1/80th of the speed
of light in vacuum, and for $\varsigma$$=$$0.1$ the minimum group velocity
is one tenth of the speed of light in vacuum.
Because of the lower overlap of the electromagnetic field with the
dielectric solid in the air bands, the effect of loss on the group
velocity is less pronounced for these bands.
While the value of the minimum group velocity is clearly highly
dependent on the conductivity, the frequency $\omega_0$ at which the
group velocity is at its minimum remains nearly constant, except for a
very slight red-shift at high values of the conductivity.
In Fig.~\ref{fig:braggVg}b we show the maximum and minimum values of the
group velocity as a function of the dimensionless conductivity.
The maximum values of the group velocity follow
very precisely the $1/\epsilon_2''$ dependence expected from our
initial analysis.
Assuming a similar distribution of the induced DOS in both band gaps,
this analysis also indicates why the maximum group velocity
falls off more rapidly in the case of the second, narrower band gap.
The minimum group velocity is found to follow almost exactly
the expected $\sqrt{\epsilon_2''}$ dependence.
We note again that the
dependence of the minimum group velocity on loss is less pronounced for the air
band, due to the smaller overlap of the electromagnetic mode with the
lossy dielectric.
We have verified that these simple scaling laws of the minimum and
maximum group velocities hold for any parameters of the Bragg stack,
and also in the case of oblique incidence.

\begin{figure}
\begin{center}
\epsfig{file=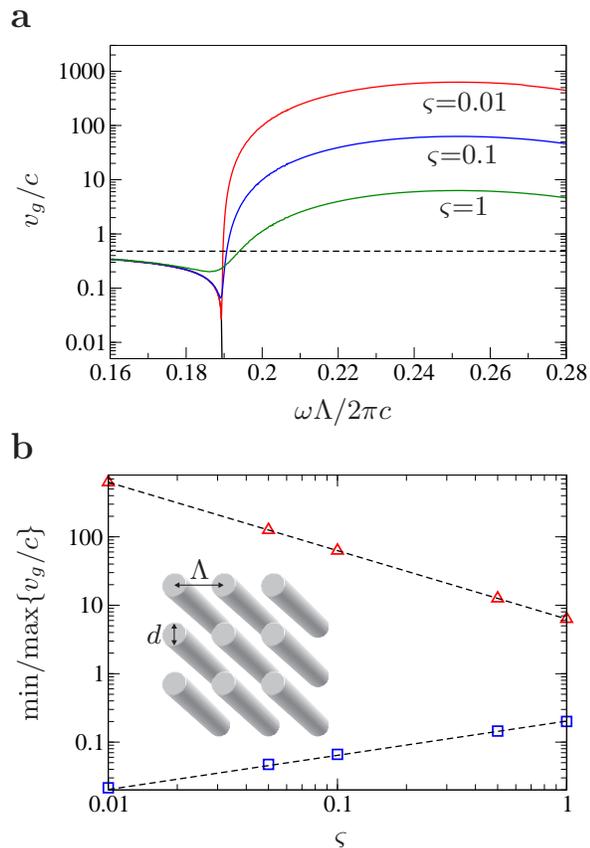, width=.9\linewidth,clip}
\caption{
(a) Group velocity as a function of
frequency for three different values of the dimensionless
parameter $\varsigma$=$\sigma\Lambda/(2\pi c\epsilon_0)$,
characterizing the conductivity of the dielectric solid.
The solid
black line corresponds to the case of zero conductivity.
The dashed horizontal
line indicates the group velocity of a homogenous material with
$\epsilon=\epsilon_1+(\epsilon_2'-\epsilon_1)\pi r^2/\Lambda^2$.
(b) The minimum and maximum group velocities
as a function of the dimensionless conductivity.
The dashed black lines indicate fits to the numerical
data of the form $A\sqrt{\varsigma}$ for the minimum group
velocities, and of the form $A/\varsigma$ for the maximum
group velocities. The inset shows the geometry of
the photonic crystal.
}
\label{fig:2DVg}
\end{center}
\end{figure}
We next consider a two-dimensional photonic crystal consisting of
dielectric cylinders of diameter $d/\Lambda=0.55$ surrounded by
air and arranged in a square lattice with lattice constant $\Lambda$.
We take $\epsilon_2'=(3.88)^2$ for the dielectric cylinders,
corresponding to silicon, and
model the loss via the Drude model, as for the Bragg stack. We
consider normal incidence of TM modes.
As for the Bragg stack, the effect of loss is to induce a density of
states in the PBGs of the structure.
The dispersion relations are calculated using the transfer matrix
method~\cite{Pendry1992}.
In Fig.~\ref{fig:2DVg}a we show the
group velocity as a function of normalized frequency for increasing
values of the dimensionless conductivity. The results shown are for
frequencies near the first TM photonic band gap. Similar trends are
seen for the 2D photonic crystal as for the Bragg stack, namely
that the density of states induced by the finite conductivity of
the dielectric resolves the divergence of the density of states
at the band gap edge, thus causing $v_g>0$ at this point. Also,
superluminal group velocities are attained inside the photonic
band gap.
In Fig.~\ref{fig:2DVg}b the minimum and maximum
group velocities are shown as functions of the dimensionless
conductivity. As for the Bragg stack, the maximum and minimum
group velocities are approximated very precisely by $1/\epsilon_2''$
and $\sqrt{\epsilon_2''}$ dependencies, respectively, supporting
our conclusion that these scalings are indeed universal for any
geometry, and even dimensionality, of the photonic crystal.

In this paper, we have for simplicity focused on conductivity as the cause of
the smearing of the density of states. However, it is clear that
any effect leading to a finite broadening
of the electromagnetic states will result in a similar scaling
of the minimum attainable group velocity with the order of the
broadening.
This includes loss, disorder, finite-size effects, etc. and
consequently imposes significant limits
on the minimum group velocities attainable via photonic crystals.
Even in the absence of other broadening mechanisms, absorption
will inevitably be present and thus represents the ultimate
limiting factor on the attainable reduction in group velocity.
In closing, we note that while we have focused
our attention to photonic crystals, the analysis presented
in this paper applies equally well to any slow-light scheme
based on band structure effects, such as for example coupled
resonator waveguides, where it
has been shown that lattice disorder may severely limit
the attainable reduction in group velocity~\cite{Mookherjea2007}.

\section{Acknowledgments}
This work is financially supported by the
\emph{Danish Council for Strategic Research} through the
\emph{Strategic Program for Young Researchers} (grant no:
2117-05-0037) as well as the \emph{Danish Research Council for
Technology and Production Sciences} (grants no: 274-07-0080 \&
274-07-0379). The work has been performed in the frame of
the Villum Kann Rasmussen Centre of Excellence "NATEC"
(Nanophotonics for Terabit Communications).

\end{document}